\documentclass[showpacs,10pt,twocolumn,prl]{revtex4-1}
\usepackage{CJK}
\usepackage{amsmath}
\usepackage{amssymb}
\usepackage{graphicx}

\begin{document}

\title{Electronic Griffiths Phase in the Te - Doped Semiconductor FeSb$_{2}$}
\author{Rongwei Hu,$^{1,\ast }$ Kefeng Wang,$^{1}$Hyejin Ryu,$^{1}$ Hechang Lei,$^{1}$ E. S. Choi,$^{2}$ M. Uhlarz,$^{3}$ J. Wosnitza,$^{3}$ and C.
Petrovic$^{1,\ddag}$}
\affiliation{$^{1}$Condensed Matter Physics and Materials Science Department, Brookhaven
National Laboratory, Upton, NY 11973\\
$^{2}$NHMFL/Physics, Florida State University, Tallahassee, Florida 32310, USA\\
$^{3}$Hochfeld-Magnetlabor Dresden (HLD), Helmholtz-Zentrum Dresden-Rossendorf, D-01314 Dresden, Germany}
\date{\today}

\begin{abstract}
We report on the emergence of an Electronic Griffiths Phase (EGP) in the doped semiconductor FeSb$_{2}$, predicted for disordered insulators with random localized moments in the vicinity of a metal-insulator transition (MIT). Magnetic, transport, and thermodynamic measurements of Fe(Sb$_{1-x}$Te$_{x}$)$_{2}$ single crystals show signatures of disorder-induced non-Fermi liquid behavior and a Wilson ratio expected for strong electronic correlations. The EGP states are found on the metallic boundary, between the insulating state ($x = 0$) and a long-range albeit weak magnetic order ($x \geq 0.075$).
\end{abstract}

\maketitle

The coupling of disorder to strong correlations remains one of the most interesting frontiers in physics \cite{Miranda}. One of the key issues is the mechanism of disorder-induced quantum Griffiths phases in metals. Magnetic quantum Griffiths phases (MGP) found in insulators host rare magnetic clusters with large susceptibilities \cite{Griffiths}. They arise in the proximity to magnetic phase transitions and have been advocated recently to be the correct framework for understanding of the non-Fermi-liquid behavior even in metallic heavy-fermion compounds \cite{Andrade,Neto2}. Yet, it was pointed out that in such systems quantum tunneling will be suppressed, leading to superparamagnetic classical instead of quantum Griffiths behavior \cite{Millis,Millis2}. The quantum Griffiths singularities can only be observed in the temperature region $T^{\ast }$ $<$ $T$ $<$ $T_{F}$, where $T_{F}$ is the Fermi temperature, whereas below $T^{\ast }$ the clusters are frozen and electronic heat capacity $\gamma \sim lnT$ and magnetic susceptibility $\chi \sim 1/T$ \cite{Neto2,Neto3}. However, the crossover temperature to the paramagnetic phase $T^{\ast }$ and the temperature window where Griffiths phases can be observed in heavy-fermion metals remain controversial \cite{Neto4,Millis3,Vojta}. On the other hand, EGP in metals are found close to a disorder-driven Mott-Anderson transition where the number of unscreened local spins rises with the increase in disorder strength \cite{Vlada,Aguiar,Tanaskovic}. Nevertheless, the Kondo or magnetic materials where EGP has been unambiguously observed remain elusive due to the stability of competing ground states such as MGP or Kondo-cluster-glass \cite{Vlada2,Guo,Kassis,Westerkamp}.

FeSb$_{2}$ is an example of a nearly magnetic (Kondo insulator-like) semiconductor, similar to FeSi \cite{Petrovic2,Perucchi,Lukoyanov}. It exhibits a large Seebeck coefficient $S$ and the highest known thermoelectric power factor $S^{2}\sigma $ \cite{BentienEPL}. The electronic system in Fe(Sb$_{1-x}$Te$_{x}$)$_{2}$ shows a high sensitivity to substitutions. An MIT is induced at the critical concentration $x_{c}=0.001$ and a region of canted antiferromagnetism is observed for $0.1\leq x\leq 0.4$ with an intermediate ferromagnetic phase for $x=0.2$ \cite{Rongwei6}. We now focus on the paramagnetic metallic region in the vicinity of the MIT for $x_{c}<x<0.1$ where we note that a canted antiferromagnetic
state is induced already at $x = 0.075$. Below that concentration, both thermodynamics and electrical transport are consistent with the EGP predicted in disordered heavy-fermion metals.

Single crystals of Fe(Sb$_{1-x}$Te$_{x}$)$_{2}$ were prepared as described previously \cite{Rongwei6}.
They were oriented using a Laue camera and polished along three principal crystal axes for four-probe resistivity measurements. Magnetic, thermal and transport measurements
were carried out in a Quantum Design MPMS-5, $^{3}$He inserts of PPMS-9, PPMS-14, and in an Oxford $^{3}$He insert equipped with an 18 Tesla magnet at NHMFL, Tallahassee.

The low-temperature heat capacity of Fe(Sb$_{1-x}$Te$_{x}$)$_{2}$ above 0.7 K is best described with a $C(T)/T=\beta
T^{2}+\delta T^{4}+cT^{-1+\lambda_{C}}$ (Fig. 1(a)). The first two terms are due to harmonic and anharmonic phonon contributions \cite{Cezairliyan}. The last term is a disorder driven non-Fermi-liqiud (NFL) electronic term with $\lambda_{C}<1$,
predicted for Griffiths phases in disordered paramagnetic metals \cite{Neto}. Fits to our data (red lines)
are excellent down to 0.7 K and the fitted parameters are given in Table I. The phonon contribution remains nearly the
same, as would be expected for such low doping level. The difference in heat capacity below 10 K stems mainly from the electronic
part. Below 0.7 K, $C(T)/T$ shows additional upturn. This cannot be
described by a paramagnon model for spin fluctuations, $C(T)/T=\gamma
_{SF}+\delta _{SF}T^{2}\ln T/T_{SF}$ \cite{Doniach}.
Moreover, the contribution due to spin fluctuations is generally sensitive to
magnetic field, i.e. such paramagnon fluctuations would be suppressed by magnetic field
and this is not observed. A power law, $C(T)/T\sim aT^{-3}$, (dashed line, Fig. 1)
represents well the low-temperature heat-capacity increase below 0.7 K, suggesting a nuclear Schottky anomaly or a low lying magnetic ground state below 0.4 K similar to YbRh$_{2}$Si$_{2}$ or U$_{x}$Y$_{1-x}$Pd$_{3}$ \cite{Trovarelli, Ott}.

\begin{figure}[th]
\centerline{\includegraphics[scale=0.7]{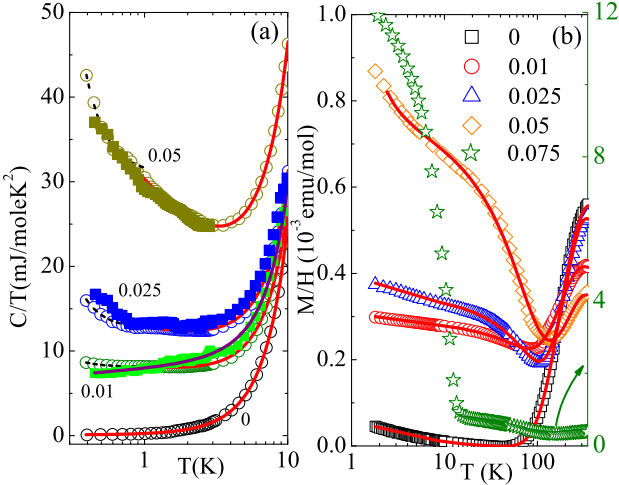}} 
\vspace*{-0.3cm}
\caption{(a) (Color online) $C{}_{p}/T$ versus $T$ for $x\leq0.05$ in Fe(Sb$_{1-x}$Te$_{x}$)$_{2}$. The $C(T)\sim T^{-2}$ dependence is represented by
the dashed lines below 0.7 K. Closed symbols show data taken in $\mu_{0}$H = 9 T. Red and purple lines are fits to the
formula given in the text. (b) Polycrystalline average of the magnetic susceptibility of Fe(Sb$_{1-x}$%
Te$_{x}$) as a function of temperature for $0\leq x\leq 0.075$ in 1 kOe magnetic field. Solid lines
are the best fits to the formula given in the text.}
\end{figure}

\begin{figure}[bh]
\centerline{\includegraphics[scale=0.65]{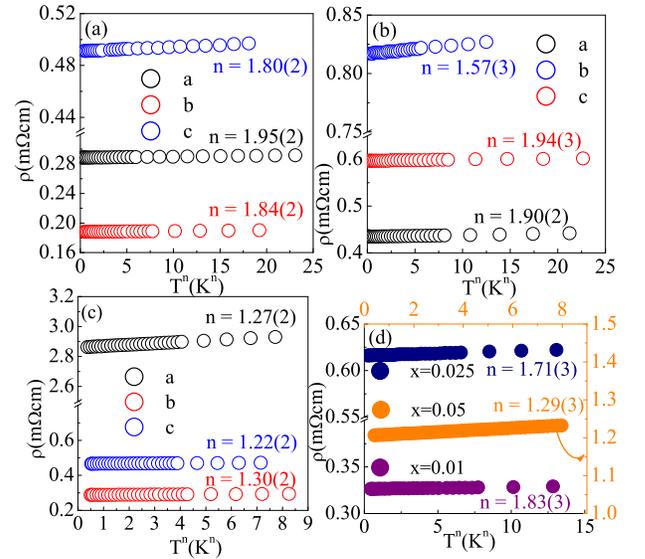}} 
\vspace*{-0.3cm}
\caption{(Color online) (a)-(c) Anisotropic $\rho$(T) for $x$=0.01, 0.025 and 0.05 in Fe(Sb$_{1-x}$Te$_{x}$)$_{2}$ (d) Polycrystalline resistivity of samples as a function of $T^{n}$ for different Te concentrations.}
\end{figure}

Deviation from the Fermi liquid behavior is often found near a magnetic instability. LDA+U calculations have revealed that FeSb$_{2}$
is in the proximity to a magnetic ground state with an on-site Coulomb repulsion $U=2.6$ eV \cite{Lukoyanov}.
As Te substitutes Sb in the lattice, the thermally activated susceptibility of FeSb$_{2}$ diminishes, accompanied by an increasing low-temperature tail (Fig. 1(b)). The $\chi(T)$ data can be
well described by
\begin{equation*}
\chi (T) = \chi _{NB}(T)+\frac{C_{1}}{T-\Theta _{1}}+aT^{-1+\lambda_{\chi}}.
\end{equation*}%
The first two terms are the narrow-band-gap \cite{Petrovic2,Rongwei6,Mandrus} susceptibility $\chi _{NB}(T)$ (as described in Ref. 22) and the impurity-related Curie-Weiss term whose Curie constant $C_{1}$ corresponds to the moment $\mu_{1}$ ($C_{1}=\mu_1^{2}/8$) and Weiss temperature $\theta_1$. As small amounts of Te enter the lattice,
the gap $\Delta$ and impurity moment related parameters $\mu_1$ and $\theta_1$ are approximately constant whereas the bandwidth $W$ increases considerably (Table I) \cite{Rongwei6}.
The additional term, $\chi(T) \sim T^{-1+\lambda_{\chi}}$ corresponds to the magnetic susceptibility expected for Griffiths phases \cite{Neto}.
At $x=0.075$, we observe signatures for magnetic order below 12 K, similar to what was earlier observed for $x=0.1$ \cite{Rongwei6}.
Fitted parameters are listed in Table I.

Above the critical concentration ($x_{c}=0.001$), the anisotropy is greatly reduced and the electrical resistivity is metallic for current applied along all 3 axes of the orthorombic unit cell (Fig. 2(a-c)) for $0.01\leq x\leq 0.05$ \cite{Rongwei6}. This is in line to isotropic $\chi (T)$ in the same temperature region. For $0.075\leq x\leq
0.2$, $\rho(T)$ is similar to $x=0.1$ and $x=0.2$ \cite{Rongwei6} and the semiconducting gaps from
fits to the activated behavior are less than 1 K, suggesting that the ground
state is semimetallic. The metallic $\rho(T)$ in the paramagnetic region ($0.01\leq
x\leq 0.05$) is characterized by a power-law temperature dependence $\rho$ $\sim$ $T^{n}$, where ($1$ $\leq$ $n$ $\leq$ $2$) for current applied along all 3 principal axes of the orthorombic structure. The exponent $n$ diminishes as more Te enters the lattice (Fig. 2(d)) and the electronic system is tuned from Kondo-insulator-like nonmagnetic towards a weak magnetic ground state. However, even a modest magnetic field $\mu$$_{0}H$ = 2 T induces a Fermi-liquid-like resistivity $\rho$ $\sim AT^{2}$ in a crystal with $x=0.01$ (Fig. 3(a)). The low-temperature resistivity is quadratic up to 18 T and we plot the coefficient $A$ as a function of field in Fig. 3(b). As opposed to the divergence in $A$ near a magnetic-field-induced quantum critical point, we observe only modest increase \cite{Gegenwart,Paglione}. Examination of $C/T$ data in $\mu$$_{0}H$ = 9 T shows a field independence for $x=0.05$ and $x=0.025$ \cite{comment}, whereas for $x=0.01$ there is also a small change in the fitting parameters $\beta$, $\delta$, and $c$: ($\beta$ = 0.12(2), $\delta$ = 0.0008(2), $c$ = 7.9(1), $\gamma_{0}$ = 7.4(2) mJ/moleK$^{2}$). However, for $x=0.01$, $\lambda$ = 1.07(2), suggesting that the electronic system has been tuned away from the NFL behavior ($\lambda_{C},\lambda_{\chi}$ $<$ 1) \cite{Tanaskovic}.

\begin{figure}[bh]
\centerline{\includegraphics[scale=0.65]{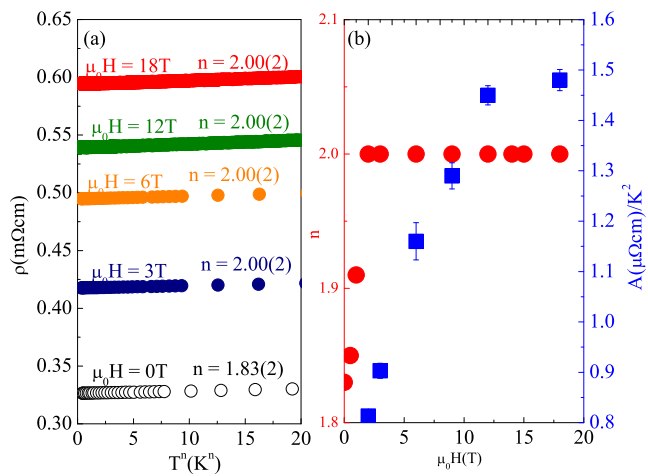}} 
\vspace*{-0.3cm}
\caption{(Color online) (a) Low temperature resistivity of Fe(Sb$_{1-x}$Te$_{x}$)$_{2}$ for $x=0.01$ plotted vs $T^{n}$ in magnetic fields up to $\mu$$_{0}H$ = 18 T. Exponent $n$ from (a) and coefficient A in $\rho$ = $\rho_{0}$+$AT^{2}$ as a function of magnetic field (d).}
\end{figure}

\bigskip
\begin{table*}[th]
\caption{Fitting parameters of the magnetic susceptibility and heat
capacity. $W$ is the bandwidth and $N(E_{F}$) is in unit of state/eV F.U. FeSb$_{2}$ has negligible $\gamma_{0}$. The impurity terms $\mu_1$ and $\theta_1$ are observed already in the pure material and remain similar in the entire investigated doping range.}%
\begin{tabular}{ccccccccccccccc}
\hline\hline
$x$ & $\Delta (K)$ & $W(K)$ & $\mu _{1}(\mu _{B})$ & $\Theta _{1}$ (K) & $a$ & $%
\lambda _{\chi }$ & $\lambda _{C}$ & $\gamma _{0}$ (mJ/molK$^{2}$) & $\beta $ & $\delta $ & $%
c$ & $N(E_{F})$ & $m^{\ast }(m_{e})$ & $R_{W}$ \\ \hline
$0$ & $425(9)$ & $310(8)$ & $0.030(2)$ & $0.8(2)$ &  &  &  & $ \sim 0$ & $%
0.16(1)$ & $0.0008(9)$ &  & $0.0039(4)$ &  &  \\
$0.01$ & $436(7)$ & $451(6)$ & $0.035(3)$ & $1.6(3)$ & $856(9)$ & $0.86(3)$
& $0.91(7)$ & $8.7(2)$ & $0.12(7)$ & $0.0007(6)$ & $8.1(8)$ & $3.7(2)$ & $%
21(1) $ & $2.7$ \\
$0.025$ & $448(4)$ & $525(9)$ & $0.036(1)$ & $3.7(3)$ & $1117(3)$ & $0.84(2)$
& $0.87(5)$ & $13.9(3)$ & $0.15(9)$ & $0.0005(9)$ & $12.8(9)$ & $5.9(3)$ & $%
25(1)$ & $2.1$ \\
$0.05$ & $453(5)$ & $525(9)$ & $0.039(2)$ & $1.8(5)$ & $1078(9)$ & $0.89(4)$
& $0.72(3)$ & $39.2(3)$ & $0.27(3)$ & $0.0002(1)$ & $30.2(8)$ & $16.7(4)$ & $%
56(2)$ & $2.3$ \\ \hline\hline
\end{tabular}%
\end{table*}

The zero-temperature electronic heat capacity $\gamma_{0}$ (Table I) was obtained by extrapolation of the curves in Fig. 1 to $T$ = 0 with the upturn below 0.7 K subtracted. The $\gamma_{0}$ values increase
monotonically from nearly zero for $x=0$ to $\gamma=39.2$ mJ/mol K$^{2}$ for $%
x=0.05$. By making a coarse estimate that the Fermi surface in Fe(Sb$%
_{1-x}$Te$_{x}$)$_{2}$ is spherical and counting one itinerant electron per Te, we may estimate the electron-mass enhancement due to electronic
correlations. The quasiparticle effective masses are moderately renormalized and reach a
maximum of 56 m$_{e}$ for $x=0.05$, reminiscent of the heavy-fermion state
induced in FeSi$_{1-x}$Al$_{x\text{ }} $(Table I) \cite{Ditusa}. This is in agreement with the Wilson ratio ($R_{W}=4\pi ^{2}k_{B}^{2}/(g3\mu _{B}^{2})(\chi
/\gamma )$) which is a standard measure of electronic correlations in metals (Table I) \cite{Wilson}. $R_{W}$ is expected to be unity for a free electron gas and it is about 2 if strong electronic correlations are present.

\begin{figure}[th]
\centerline{\includegraphics[scale=0.7]{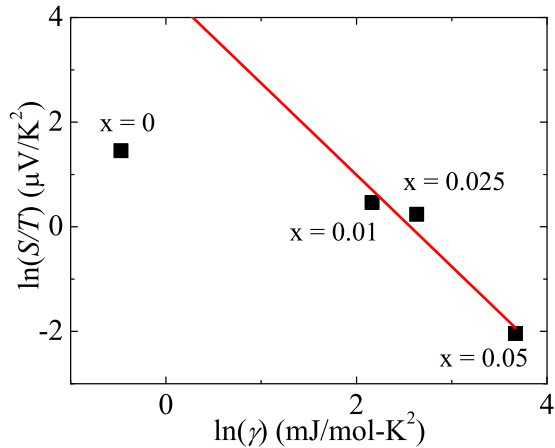}} 
\vspace*{-0.3cm}
\caption{(Color online) Power-law dependence of $\frac{S}{T}$ on $\gamma$ for Fe(Sb$_{1-x}$Te$_x$)$_2$. The solid line is the fit to the data (excluding $x=0$ point with an exponent of -1.7(4). The direction of heat current transport was along the crystallographic $c$-axis.}
\end{figure}

In metals conductivity arises due to electron diffusion and the Seebeck coefficient can be described as $S=\frac{\pi^2k_B^2}{2e}\frac{T}{T_F}=\frac{\gamma T}{ne}$, where $\gamma$ is given by $\gamma=\frac{C_{el}}{T}=\frac{\pi^2}{2}\frac{nk_B}{T_F}$, with $C_{el}$ the electronic specific heat and $n$ the carrier density \cite{Pourret}. The correlation between $\frac{S}{T}$ and $\gamma$ in the zero-temperature limit gives a scaling in metals as $q=\frac{SN_{av}e}{T\gamma}=\pm 1$ where $N_{av}$ is Avogadro's constant \cite{Behnia}. This quasi-universal ratio is valid if the relevant scattering arises from impurities but not from the mutual scattering of quasiparticles \cite{Miyake}. In order to shed more light onto the nature of correlations in our crystals, we performed Seebeck-coefficient measurements at $\sim 2$ K. Fig. 4 shows the power-law dependence of $\frac{S}{T}$ on $\gamma$ for Fe(Sb$_{1-x}$Te$_x$)$_2$ with $x = 0, 0.01, 0.025$, and $0.05$. The solid line is the power-law fit to the data for metallic samples (excluding $x=0$). FeSb$_2$ ($x=0$) is a Kondo-insulator-like narrow-gap semiconductor with extremely low carrier density. Hence, a very large magnitude of $q\sim 663$ corresponding to a small but finite carrier density of 0.0015 electron per unit cell ($\sim$ $1.25$ $\times$ $10^{19}$ cm$^{-3}$) is obtained. Similar effect was discovered in CeNiSn \cite{Behnia}. The doped samples exhibit a power-law dependence of $S/T$ on $\gamma$ with an exponent $\sim$$-1.7$$\pm$$0.4$ (Fig. 4), contradicting the quasi-universal ratio in a Fermi liquid and implying very strong
electron-electron scattering and intersite magnetic correlations. The power-law dependence of $S/T$ on $\gamma$ is expected for metals near a MIT, as seen in FeSi$_{1-x}$Al$_{x}$ \cite{Sharath}. This shows that the proximity to a MIT has substantial effects on thermodynamics and, as we discuss below, on the mechanism of the Griffiths phase.

The increase of disorder from a clean insulator will create states in the gap in the EGP scenario, increasing the density of states at the Fermi level until the gap is decreased to zero at the MIT, giving rise to metallic states with disordered local moments \cite{Aguiar}. Even more disorder tunes the system again to an insulating state due to localization effects \cite{Vlada,Eduardo}. Whereas the EGP formalism has been developed for disordered Kondo systems, we note that the Hubbard $U$ in the strongly correlated electronic system of is FeSb$_{2}$ is reduced from the values common in lanthanide based heavy fermions and Fe moments are not as localized \cite{Tomczak,Igor,Sun}. Nevertheless, the EGP predictions are in qualitative agreement with the Fe(Sb$_{1-x}$Te$_{x}$)$_{2}$ phase diagram \cite{Rongwei6} and with the values we obtained in Table I. Metallic conductivity with a characteristic non-Fermi-liquid temperature dependence for $x = 0.01, 0.025$, and $0.05$ arises due to a bandwidth increase with Te substitution. The non-Fermi liquid exponent \textit{n} in $\rho(T) \sim T^{n}$ (Fig. 2) is close to \textit{n=3/2}, expected both in metallic glasses \cite{Jaroszynski} as well as for disordered heavy-fermion metals near an antiferromagnetic quantum critical point (QCP) \cite{Rosch}. We note that the resistivity should evolve from $\rho \sim T^{1}$ to $\rho \sim T^{3/2}$ as the disorder is increased near a magnetic QCP \cite{Rosch}, in contrast to the decreasing $n$ with $x$ in Fe(Sb$_{1-x}$Te$_{x}$)$_{2}$ (Fig. 2(d)). In addition, large values of the Wilson ratio (R$_{W}$ $\sim$ 20 - 700) are expected in MGP or in cluster-glass phases due to the existence of large-moment magnetically ordered clusters with a Griffiths exponent $\lambda \rightarrow 0$ as the system is tuned to the magnetic phase \cite{Vojta,Qu,Westerkamp}. This is in contrast to our observations where R$_{W}$ $\sim$ 2 and $\lambda \sim (0.8-0.9)$ (Table I). This discussion suggests a considerable influence of the charge channel (MIT) in the underlying microscopic mechanism of the Griffiths phase. Spin fluctuations near a glassy state in the EGP are expected to lead to a marginal Fermi-liquid behavior \cite{Tanaskovic2} and a nearly logarithmic divergence of $\gamma$ and $\chi$ ($\lambda$ $\rightarrow$ $1$), all in much closer agreement with the parameters obtained here (Table I).

In summary, we have shown that a small amount of Te doping in Fe(Sb$%
_{1-x}$Te$_{x}$)$_{2}$ single-crystal alloys (up to $x\leq 0.05$) results in the non-Fermi liquid Griffiths metallic state that primarily has an electronic origin. This highlights the importance of fluctuations in the conduction-electron local density of states. The EGP theoretical framework rests on the power-law distribution of the energy scale for the spin fluctuations, not necessarily of Kondo origin \cite{Miranda}. However, similarity with disordered Kondo/Anderson lattices is still significant since NFL divergences appear at rather weak randomness ($x=0.01$) and are magnetic field tuned to a Fermi liquid.

We thank Vladimir Dobrosavljevic for useful communication. Work at Brookhaven is supported by the U.S. DOE under
Contract No. DE-AC02-98CH10886. Work at the National High Magnetic Field Laboratory is supported by NSF Cooperative Agreement No. DMR-0084173, by the State of
Florida, and by the U.S. Department of Energy. Part of this work was supported by EuroMagNET II (EU Contract No. 228043). CP acknowledges support by the Alexander von Humboldt Foundation.
$\\$

$^{\ast }$Present address: Department of Physics, University of Maryland, College
Park, MD 20742-4111, USA.

${\ddag }$ petrovic@bnl.gov


\begin{thebibliography}{99}
\bibitem{comment} We have evidence that the resistivity coefficient $n$ also increases in magnetic field for $x=0.025$ and $x=0.05$ but never reaches $n$ = 2 up to $\mu$$_{0}H$ = 18 T.
\end{thebibliography}
\end{document}